\documentclass[preprint,sort&compress,  5p]{elsarticle}

\pdfpageattr{/Group << /S /Transparency /I true /CS /DeviceRGB >>}

\usepackage[utf8]{inputenc}
\usepackage[T1]{fontenc}
\usepackage{lmodern}
\usepackage{textcomp}
\usepackage{fixltx2e}
\usepackage{microtype}
\usepackage{calc}
\usepackage{multirow}
\usepackage{multicol}

\usepackage{amsmath}
\usepackage{amssymb}
\usepackage{amsfonts}
\usepackage{amsthm}

\usepackage{booktabs}
\usepackage{url}

\usepackage[caption=false,font=footnotesize]{subfig}
\usepackage{graphicx}
\DeclareGraphicsExtensions{.pdf,.png,.jpg}
\usepackage{tikz}
\usetikzlibrary{calc}

\usepackage[american]{babel}
\usepackage{hyphenat}

\usepackage{algorithm}
\usepackage{algpseudocode}

\usepackage[capitalize,noabbrev]{cleveref}

\crefname{algocf}{Algorithm}{Algorithms}
\Crefname{algocf}{Algorithm}{Algorithms}

\usepackage{cases}
\usepackage[version=4]{mhchem} 

\begin{document}

\begin{frontmatter}
		
	\title{Mobile Human Ad Hoc Networks: A Communication Engineering Viewpoint on Interhuman Airborne Pathogen Transmission}

		\author[address1,address2]{Fatih Gulec\corref{cor1}}
		\ead{fgulec@yorku.ca}
		\cortext[cor1]{Corresponding author (Current address: Department of Electrical Engineering and Computer Science, York University, Toronto, Ontario, Canada)}
		
		\author[address2]{Baris Atakan}
		\ead{barisatakan@iyte.edu.tr}
		
		\author[address1]{Falko Dressler}
		\ead{dressler@ccs-labs.org}
		
		\address[address1]{School of Electrical Engineering and Computer Science, TU Berlin, Germany}		
		\address[address2]{Izmir Institute of Technology, Department of Electrical and Electronics Engineering, Izmir, Turkey}

\journal{Nano Communication Networks}

	\begin{abstract}
A number of transmission models for airborne pathogens transmission, as required to understand airborne infectious diseases such as COVID-19, have been proposed independently from each other, at different scales, and by researchers from various disciplines.
We propose a communication engineering approach that blends different disciplines such as epidemiology, biology, medicine, and fluid dynamics.
The aim is to present a unified framework using communication engineering, and to highlight future research directions for modeling the spread of infectious diseases through airborne transmission.
We introduce the concept of mobile human ad hoc networks (MoHANETs), which exploits the similarity of airborne transmission-driven human groups with mobile ad hoc networks and uses molecular communication as the enabling paradigm.
In the MoHANET architecture, a layered structure is employed where the infectious human emitting pathogen-laden droplets and the exposed human to these droplets are considered as the transmitter and receiver, respectively.
Our proof-of-concept results, which we validated using empirical COVID-19 data, clearly demonstrate the ability of our MoHANET architecture to predict the dynamics of infectious diseases by considering the propagation of pathogen-laden droplets, their reception and mobility of humans.
	\end{abstract}
		
		
\end{frontmatter}

	\section{Introduction}
	Throughout the history, epidemics caused by infectious diseases have been a major threat to human life. Epidemic diseases such as smallpox, Spanish flu and recent COVID-19 gave rise to millions of human deaths \cite{brauer2019mathematical}. In addition, epidemics can induce mental disorders in humans and recessions in the world economy due to prevention and control measures such as lockdown. Owing to these facts, it is essential to accurately model the spread of infectious diseases among humans.
	
	The interhuman spread of infectious diseases occur via direct contact and airborne transmission\footnote{ Here, \textit{transmission} is employed synonymously with \textit{contagion} rather than its usage in communication engineering.} where pathogens are transferred from an infectious human to a susceptible one. In airborne transmission, these pathogens (viruses, bacteria, fungi, and so on) are carried by large droplets and aerosols (droplet nuclei) which are emitted via breathing, speaking, coughing and sneezing \cite{bourouiba2021fluid, mittal2020flow}. Throughout this paper, we use the term \textit{droplet} to refer to both large droplets and aerosols together.
	
	As for the modeling of infectious diseases between two humans, a human emitting expiratory droplets is an information source \cite{khalid2019communication, khalid2020modeling, amin2021viral}. When these emitted information carrying droplets are received by another human through sensory organs, we can consider there exists a communication path between them. Hence, a molecular communication (MC) perspective, has recently been proposed for airborne transmission modeling \cite{gulec2021molecular, schurwanz2021duality, schurwanz2021infectious}. \cite{schurwanz2021duality} and \cite{schurwanz2021infectious} lay the theoretical and experimental foundations of dualities between pathogen-laden droplet propagation and MC. \cite{barros2021molecular} gives a detailed review and discussion about the usage of MC for viral infection research. In \cite{gulec2021molecular}, an end-to-end system model for airborne pathogen transmission between two humans is proposed by considering this transmission as an information transfer through an air-based channel. In addition, the effect of face masks on the infection probability is investigated in \cite{lotter2021statistical}. In \cite{koca2021molecular} and \cite{pal2021vivid}, MC is employed to model the transmission of pathogens through the human respiratory tract. In these recent studies with the MC perspective, the proposed models are for static humans. However, as people displace, there exist dynamic human groups exchanging pathogens among each other. Due to their mobility, humans form different groups in an ad hoc fashion as their smart phones do in a mobile telecommunication network. Hence, an analogy between human groups and mobile telecommunication networks can be established, since they both possess an intermittent connectivity.
	
	On the other hand, researchers from many disciplines work separately at different scales to reveal the mechanisms of airborne pathogen transmission and model the behavior of epidemics. In fluid dynamics literature, researchers focus on the propagation of pathogen-laden droplets and their interactions with air \cite{bourouiba2021fluid, mittal2020flow}. Biologists deal with the survival of airborne pathogens in macroscale \cite{schaffer1976survival} and their interactions with the human cells in microscale \cite{cohen2016viruses}. Furthermore, the medical literature conducts researches in cellular level to discover new drugs which cure the infectious diseases. At a larger scale, epidemiology literature focuses on empirical data to develop mathematical models for the spread of epidemics in time and space \cite{brauer2019mathematical, rock2014dynamics}. However, these epidemiological models do not consider the information from fluid dynamics, biology and medicine and make estimations by fitting statistical data \cite{martcheva2015introduction}. The fluid dynamics of droplets, the geometries and air distribution of indoor environments, the pathogen-human interaction, the medical efficacy of the drugs and locations of mobile humans are essential to be taken into account for accurate models. Thus, there is a need to merge all of these research efforts in a unified framework. 
	
	To integrate the research attempts from different disciplines and utilize the analogy between mobile telecommunication networks and human groups, we propose a framework for modeling interhuman airborne pathogen transmission with communication engineering perspective. Here, mobile humans forming a group are considered as a mobile human ad hoc network (MoHANET). In a MoHANET, the infectious human is the transmitter (TX), the susceptible human is the receiver (RX) and pathogen-laden droplets are information carriers propagating in the communication channel, i.e., air. Here, MC employing chemical signals instead of electrical signals emerges as an enabler paradigm for the communication among humans due to its biocompatibility with the human body and multiscale applicability.

	Furthermore, communication engineering approach provides a unified framework by combining micro- and macro\-scale modeling issues. With this framework, a MoHANET is partitioned into layers where each layer is associated with a research area at different scales such as fluid dynamics, biology, medicine or epidemiology. As in the conventional telecommunication networks, each layer sends its outputs to an upper layer. In this way, the spread of infectious diseases can be modeled more accurately by considering all parameters from various disciplines. In addition, researchers will be able to utilize theoretical tools of communication theory in order to model the complicated nature of airborne pathogen transmission. In this paper, a proof-of-concept study is given for the MoHANET architecture. To this end, an omnidirectional multicast transmission (OMT) algorithm  in which the average number of contacts in a MoHANET is calculated by exploiting the truncated Lévy walk for the mobility of humans is proposed. The infection states of humans are determined according to their relative distance. Then, the effective contact rate is determined by using the probability of infection for airborne pathogen transmission and the average number of contacts. Lastly, this effective contact rate is employed in an epidemiological model to estimate the time course of an epidemic. Numerical results validated by empirical COVID-19 data show that the number of infected people during an epidemic can be estimated  by taking into account the propagation and reception of pathogen-laden droplets and the mobility of humans. Moreover, the results show that the increment of the population's immune system strength or the reduction in the number of received airborne pathogen-laden droplets leads to a milder outbreak over time.
	
	
	

	In the remainder of this paper, we first review the airborne pathogen transmission mechanisms. Then, the communication engineering approach which merges different disciplines is introduced. In this approach, the layered architecture of the MoHANET is presented in detail and open research issues are discussed. In the next section, proof-of-concept study is given with the proposed OMT algorithm and numerical results. Finally, we give a discussion on the existing and possible experimental techniques and conclude the paper.

	\section{Overview of Main Issues on Airborne Pathogen Transmission} \label{Overview}
\begin{figure*}[t]
	\centering
	\includegraphics[width=0.96\textwidth]{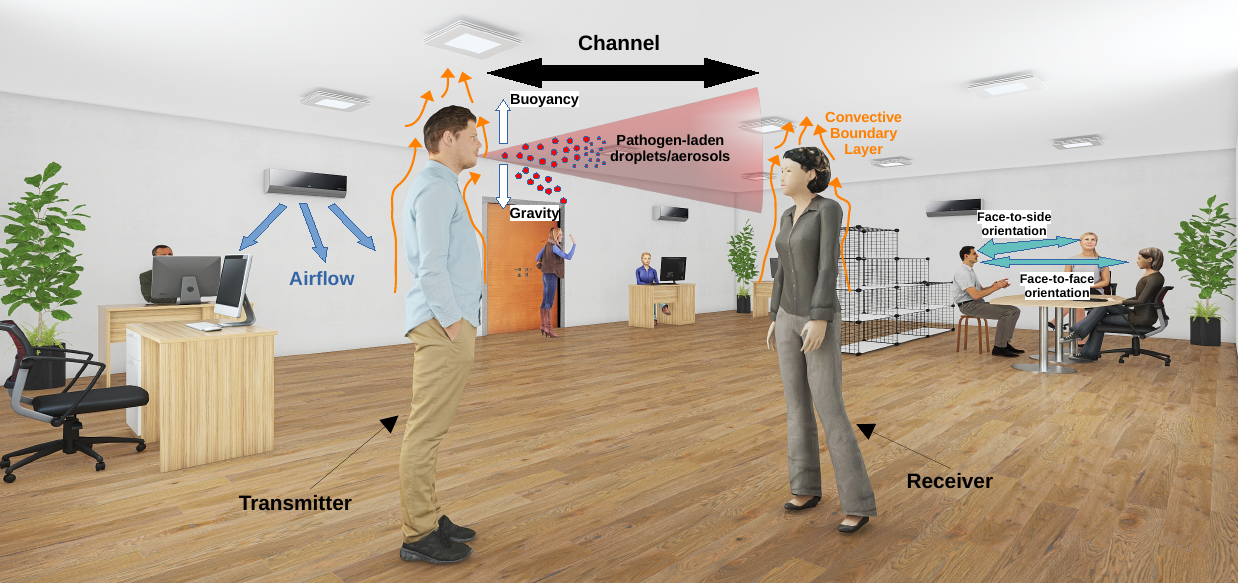}  
	\caption{
		The spread of an infectious disease through airborne pathogen transmission with communication engineering perspective and effective issues for an indoor sneezing/coughing scenario.
	}
	\label{Indoor}
	\end{figure*}

	This section provides a brief overview for the main issues of the airborne pathogen transmission mechanisms as illustrated in Fig. \ref{Indoor}. 
	\subsection{Respiratory Activity, Droplet Size and Evaporation}
	Pathogen-laden droplets are emitted to the air from an infected human via respiratory activities such as coughing, sneezing, speaking and breathing. These activities have different initial droplet velocities allowing different propagation distances. For instance, the initial velocities for coughing and breathing are about $10$ m/s and $2.67$ m/s, respectively  \cite{ai2018airborne}. Therefore, a cough can infect people at a greater distance than breathing in still air in a short time interval. However, breathing can be more effective than coughing or sneezing for longer intervals due to the continuous emission. Furthermore, the expiratory droplets are defined according to their diameters where aerosols and large droplets are assumed to have smaller and larger diameters than $10$ $\mu$m, respectively \cite{mittal2020flow}. While speaking, sneezing, and coughing release more large droplets into the air, breathing mostly contains aerosols. In addition, larger droplets settle to the ground due to gravity before evaporation and smaller droplets can become aerosols via evaporation depending on the temperature and relative humidity (RH) \cite{seminara2020biological}. For long durations, aerosols can be more infectious than large droplets, since they can remain suspended in the air and be drifted by airflows. \vspace{-0.1cm}
	\subsection{Air Distribution}
	In addition to the initial velocity, emitted droplets are influenced by the airflows, similar to a MC channel with drift. In outdoor environments, winds carry the droplets and dilute the concentration of pathogens via dispersion. Therefore, it is less probable to get infected in outdoor environments. However, in indoor environments such as hospitals or offices, airflows generated by ventilation systems are critical for the spread of pathogens due to the circulation of air in bounded conditions  \cite{ai2018airborne}. 
	
	\subsection{Posture, Relative Orientation, Distance and Movement of the Human}
	For short distances, the posture, that is, standing, sitting or lying position, and the relative orientation of the infected and susceptible persons are important for the infection risk as shown in Fig. \ref{Indoor}. For instance, a doctor can reduce the exposure from an infected lying patient in a hospital ward via a standing posture and sideways orientation instead of face-to-face orientation \cite{ai2018airborne}. Furthermore, a walking person can increase the infection risk in a closed and ventilated room by increasing the dispersion of droplets \cite{ai2018airborne}. Another important factor influencing the infection risk is the relative distance of the humans which is also referred as the social distance. Surely, the infection risk decreases, as the relative distance between two people increases. 

	\subsection{Thermofluid Boundary Conditions}
	The temperature difference between the human body surface and the surrounding air generates a thermal plume which is a buoyancy-driven upward flow of the surrounding air. As illustrated in Fig. \ref{Indoor}, this thermal plume leads to a convective boundary layer (CBL) around the human body, which should be taken into account for the movement of the droplets in the breathing zone \cite{licina2015human}. This upward flow can change the channel impulse response, which mathematically characterizes the alteration caused by the channel located between the TX and RX during airborne transmission, via generating an upward drift for the pathogens during the reception into the human body as shown in Figure \ref{Indoor}.
	\subsection{Survival of Pathogens}
	Subsequent to a respiratory activity, all of the emitted pathogens may not survive. In \cite{schaffer1976survival}, it is shown that more than $80$ percent of the influenza viruses cannot survive within one minute. However, these survival rates are severely influenced by environmental factors such as temperature and RH. While increasing temperature decreases survival rates of the pathogens due to its effect at molecular levels, increasing RH results in decreasing evaporation of droplets \cite{marr2019mechanistic}. The decreasing number of pathogens results in a time-varying channel due to the dependence on the previous number of pathogens.


	\begin{figure*}
		\centering
		\includegraphics[width=0.8\textwidth]{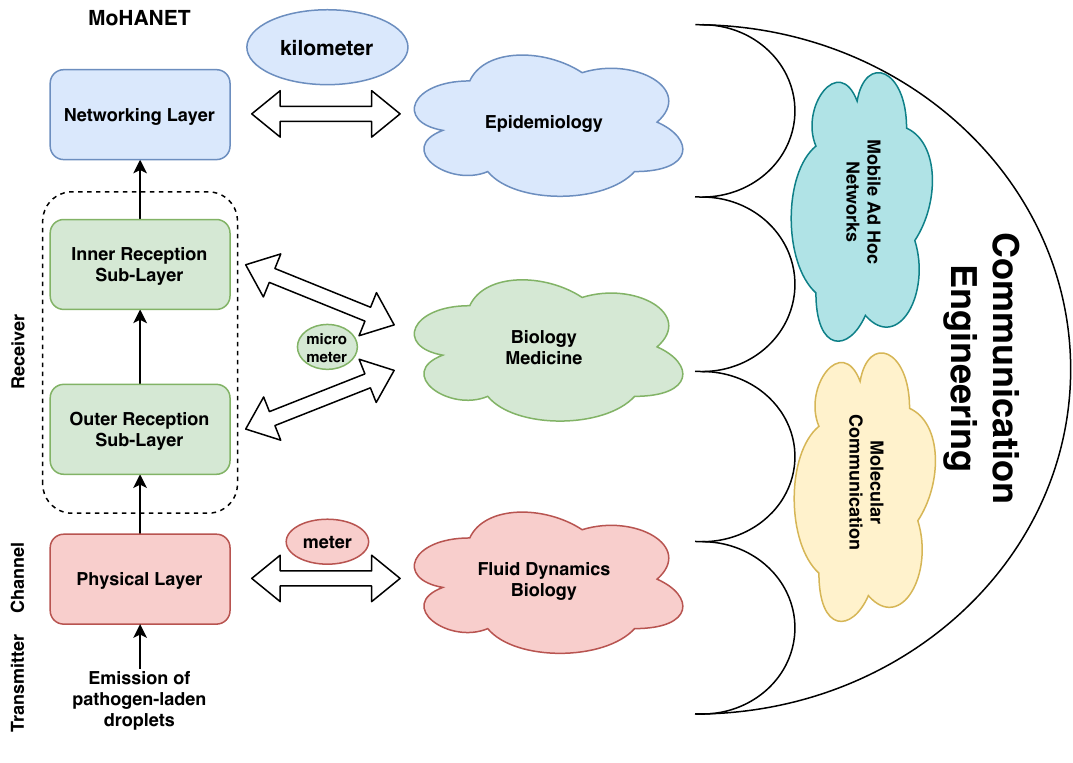}  
		\caption{
			Communication engineering framework to model the spread of infectious diseases through airborne pathogen transmission and the layered MoHANET architecture.
		}
		\label{Layer}
	\end{figure*}

	\section{Communication Engineering Approach to Interhuman Airborne Pathogen Transmission} \label{CE}
	In this section, we present a framework with communication engineering perspective to model the spread of infectious diseases through airborne pathogen transmission. Furthermore, open research issues are given. 
	
	As shown in Fig. \ref{Layer}, the proposed framework  merges all of the multiscale research efforts in various disciplines such as fluid dynamics, biology, medicine, and epidemiology under the umbrella of communication engineering. MC emerges as the key paradigm that connects the studies among different disciplines  in macro- and microscales. First, the MoHANET is introduced through a layered architecture as depicted in Fig. \ref{Layer}. Layers are associated with different disciplines from $\mu$m to km scale in this architecture where each layer sends its output to an upper layer. The first layer is defined as the physical layer where the infectious human (TX) emits pathogen-laden droplets through the communication channel (air) as illustrated in Fig. \ref{Indoor}. The next layer is the reception layer which takes place at the susceptible human (RX) and includes two sub-layers, that is, outer and inner reception sub-layers. The outer reception sub-layer comprises the interactions of the facial sensory organs with the droplets and inner reception sub-layer provides the details about the interactions of pathogens with the biological cells in the human body. The networking layer where infectious diseases spread among different people is given at the top of the MoHANET architecture where methods from mobile telecommunication networks literature and epidemiology are exploited and the outputs of the lower layers are employed. 
	
	Here, we note that there is not a one-to-one correspondence between the layers of a MoHANET and the layers of a conventional telecommunication (or computer) network. The physical layer is defined in a slightly different way from the physical layer of a conventional network, since the reception of molecular signals is not included in the physical layer of the MoHANET and information is transferred via pathogen-laden droplets instead of electromagnetic waves. The usage of the proposed layered architecture of the MoHANET is utilized to understand the phenomenon of  infectious disease spread instead of designing an efficient, high data rate and reliable telecommunication network. The details of this layered architecture are introduced as follows.
	\subsection{Physical Layer}
	\subsubsection{Transmitter}
	In a MoHANET, an infected person is considered as a TX and her/his respiratory activities determine the TX parameters such as initial droplet velocities and droplet size distribution \cite{khalid2019communication}. The respiratory activities which are mentioned earlier can be classified as impulsive (sneezing and coughing) and continuous (breathing and speaking) emission signals.  For continuous emissions, the respiration rate is an influential factor for the transmission models. In addition, the respiratory organs such as nose or mouth affect the direction of the emitted signals. For example, the infection risk increases, when the TX uses the mouth instead of nose \cite{ai2018airborne}. Furthermore, the convective boundary layer (CBL) of the human body, posture and relative orientation should be taken into account for accurate TX models. In addition, the load of pathogens in an exhaled breath or cough/sneeze, which can change according to the droplet size, temperature and RH, can affect the transmission of a disease, i.e., infectivity \cite{seminara2020biological}. On the other hand, the infectivity of a pathogen can increase during its evolution in an epidemic and can bring about super spreading events \cite{wang2020inference}. In this work, the transmitted pathogens are assumed identical and the inclusion of this infectivity, which can be estimated via a phylogenetic analysis, droplet size and environmental factors, into the model is left as an open research issue.
	
	
	
	\subsubsection{Channel}
	The channel is the physical medium between the TX and RX including the boundary conditions. As shown in Fig. \ref{Layer}, channel modeling in the physical layer requires knowledge from fluid dynamics and biology due to the air-droplet interaction and survival of pathogens, respectively. The propagation dynamics of droplets can be examined under two subheadings depending on whether there is an external airflow or not.
	\paragraph{Still Air}
	In indoor environments such as residential buildings, it is generally assumed that there is no airflow, if there is not any ventilation system. After the emission of pathogen-laden droplets with an initial velocity, they are subject to Newtonian mechanics during their interactions with the air. Emitted droplets can be modeled as a cloud consisting of droplets and air particles \cite{bourouiba2014violent, gulec2021molecular}. The movement of this cloud can be defined as a two-phase flow where these phases represent the gaseous state of air and liquid state of droplets \cite{gulec2020droplet, gulec2021fluid}. Due to gravity, large droplets may fall earlier to the ground with respect to aerosols and evaporation can shrink the size of droplets. As mentioned earlier, the temperature of air and evaporation influence the survival rates of pathogens. For continuous emissions, this fact can affect the channel memory, which is crucial for channel modeling. Furthermore, initial velocities of droplets determined by respiratory activities can give rise to laminar and turbulent flows which fade out as the distance between the TX and RX increases \cite{pendar2020numerical}.
	
	\paragraph{Windy Air}
	For windy outdoor environments and indoor environments with airflows such as ventilation or wind arising from the open doors and windows, airflows dominate the propagation of droplets rather than other factors given for still air environments. The airflow which carries the pathogen-laden droplets can be examined by advection and dispersion (turbulent diffusion) mechanisms. Briefly, advection results from the airflow velocity and dispersion depends on the turbulent eddies during the mass transfer \cite{de2013air}. It should be noted that molecular diffusion related with the thermal energy of molecules is negligible in macroscale. In order to calculate the concentration of droplets in time and space, deterministic and stochastic approaches which are based on differential Navier-Stokes and continuity equations are employed. In addition, indoor ventilation types such as under floor air distribution, mixing, displacement, and downward ventilation should be incorporated into these airflow models. For example, downward ventilation can reduce the infection risk by diluting the dispersion of droplets \cite{ai2018airborne}.
	
	\subsection{Reception Layer} \label{RL}
\begin{figure}
		\centering
		\includegraphics[width=\columnwidth]{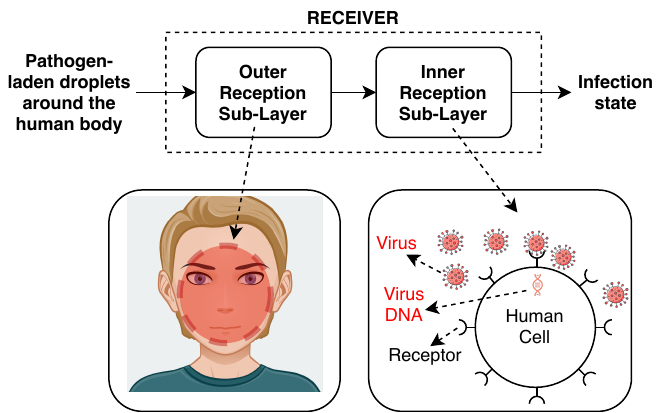}  
		\caption{
			Two-layered receiver.
		}
		\label{RX}
	\end{figure}

	A human gets infected, when the transmitted pathogens are received into the body. As shown in Fig. \ref{Layer}, the reception layer covers the issues related to biology and medicine in microscale where MC is utilized for the interactions of pathogens with human body. The reception of these pathogens by the exposed human (RX) have not been well investigated, although there are myriads of theoretical, experimental and clinical studies for the propagation of pathogens. To this end, we propose a two-layered RX as shown in Fig. \ref{RX} and detailed below.

	\subsubsection{Outer Reception Layer}
	The reception of pathogen-laden droplets occur in the eyes, mouth and nose for many pathogens such as influenza virus \cite{seminara2020biological}. Hence, we define the first step of reception as the outer layer sensing for the reception via facial sensory organs as illustrated in Fig. \ref{RX}. The whole surface of the human face is also important for the reception, since an infection may occur by touching the face contaminated with pathogens and these organs consecutively.
	
\begin{figure}
		\centering
		\includegraphics[width=\columnwidth]{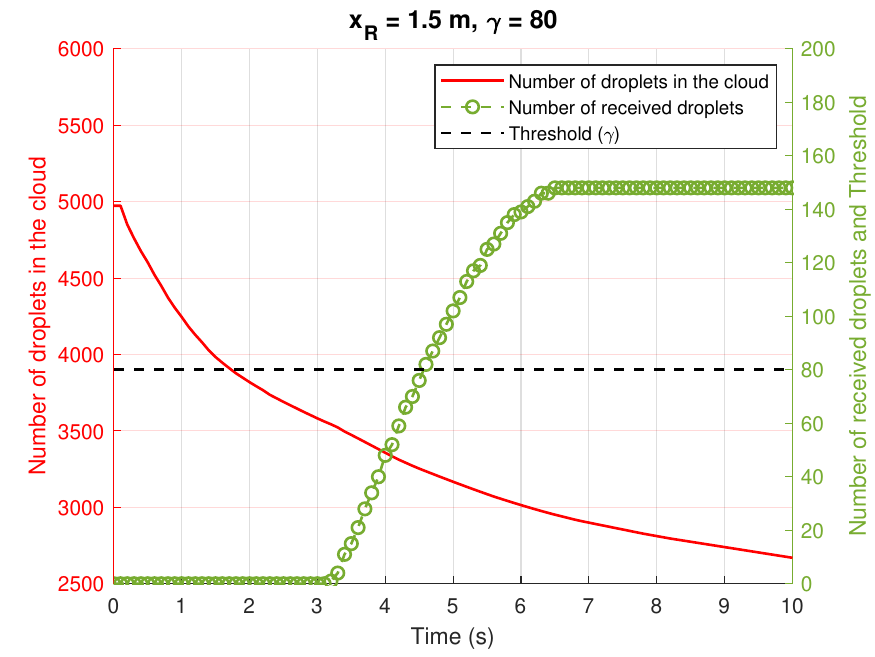} 
		\caption{
			The mean number of droplets in the cloud and their reception by the RX.
		}
		\label{Traj_R2}
	\end{figure}

	Pathogen-laden droplets emitted via a respiratory activity propagate as a mixture of droplets and air particles, which can be represented as a spherical cloud \cite{gulec2021molecular, bourouiba2014violent}. This cloud is affected by the momentum due to the initial velocity of droplets, gravity and buoyancy stemming from the temperature difference of the mouth and ambient air. According to the model detailed in \cite{gulec2021molecular}, Fig. \ref{Traj_R2} gives the change of the number of droplets in the cloud by taking settling and reception of droplets into account for a coughing TX in still air as illustrated in Fig. \ref{Indoor}. The cross-section of the RX is assumed to cover a circular area including eyes, mouth and nose at the outer layer as shown in Fig. \ref{RX}. At this point, an analogy with the communication systems can be established by considering the \textit{infected state} of the RX as symbol \textit{1} and \textit{no infection} as symbol \textit{0}. This reception is accomplished by a detection according to a threshold value ($\gamma = 80$) indicating the number of droplets required to become infected, as given in Fig. \ref{Traj_R2}. $\gamma$ is a critical parameter, since it depends on the strength of human's immune system. To this end, biomedical data of humans such as body mass index, glucose level and whether or not having chronic diseases  can be employed to estimate $\gamma$. Moreover, $\gamma$ can be effective to determine the number of infected people in an epidemic as given in Section \ref{NL}.
	
	In addition to these issues in the outer layer, the posture, relative orientation and CBL should be considered for an accurate RX model as considered for the TX. Furthermore, the reception of pathogen-laden droplets at the outer layer with different types of masks is an open issue to be investigated.
	\subsubsection{Inner Reception Layer}
	As shown in Fig. \ref{RX}, pathogens actually \textit{enter} human body at the cellular level and increase their population. For example, viruses replicate themselves by inserting their genetic material (DNA or RNA) into human cells in two ways: They can bind their fusion (or spike) protein on specific receptor sites on the human cell or they can enter by using endosomes like a Trojan horse \cite{cohen2016viruses}. Their binding sites can have different concentrations in different parts of the body. For instance, severe acute respiratory syndrome coronavirus-2, which causes COVID-19, binds to angiotensin converting enzyme-2  receptors which are mostly found at upper respiratory tract \cite{zhou2020pneumonia}. While large droplets are effective in upper respiratory tract, aerosols can reach down to alveoli in lower respiratory tract. Hence, the droplet size can be effective to determine the infection risk according to the type of the disease. Moreover, the viruses diffuse among human cells, bind to receptors and copy their genetic material in a random way. All of these issues at the inter- and intracellular level need to be modeled for an accurate transmission model for the spread of infectious diseases in MoHANETs. These modeling efforts can also contribute to drug and vaccine developments. By using the models at physical and reception layers, the infection rate can be derived to be used in the networking layer as given in the next part.

	\subsection{Networking Layer} \label{NL}
	What we examine up to here in lower layers of the MoHANET architecture is about the transmission of infectious diseases between two humans. However, these transmissions occur many times in an epidemic, which requires a perspective to handle the population as a connected group, that is, a network. In the networking layer, the details of the MoHANET architecture are presented in order to model the spread of infectious diseases at a large scale (km) within the communication engineering framework as shown in Fig. \ref{Layer}.
	
	For indoor airborne transmission, various versions of Wells-Riley model are widely employed in the literature. These models quantify the average number of pathogens by using exposure time, pulmonary and room ventilation rates in a well-mixed room with pathogens  \cite{seminara2020biological}. Although these models are useful to determine the guidelines for indoor places in an epidemic \cite{bazant2021guideline}, they lack the capability to estimate the number of infected humans over a long term for larger areas. Therefore, a different approach is adopted for long term estimations of an epidemic, which is detailed as follows.
	
\begin{figure*}
	\centering
	\includegraphics[width=1\textwidth]{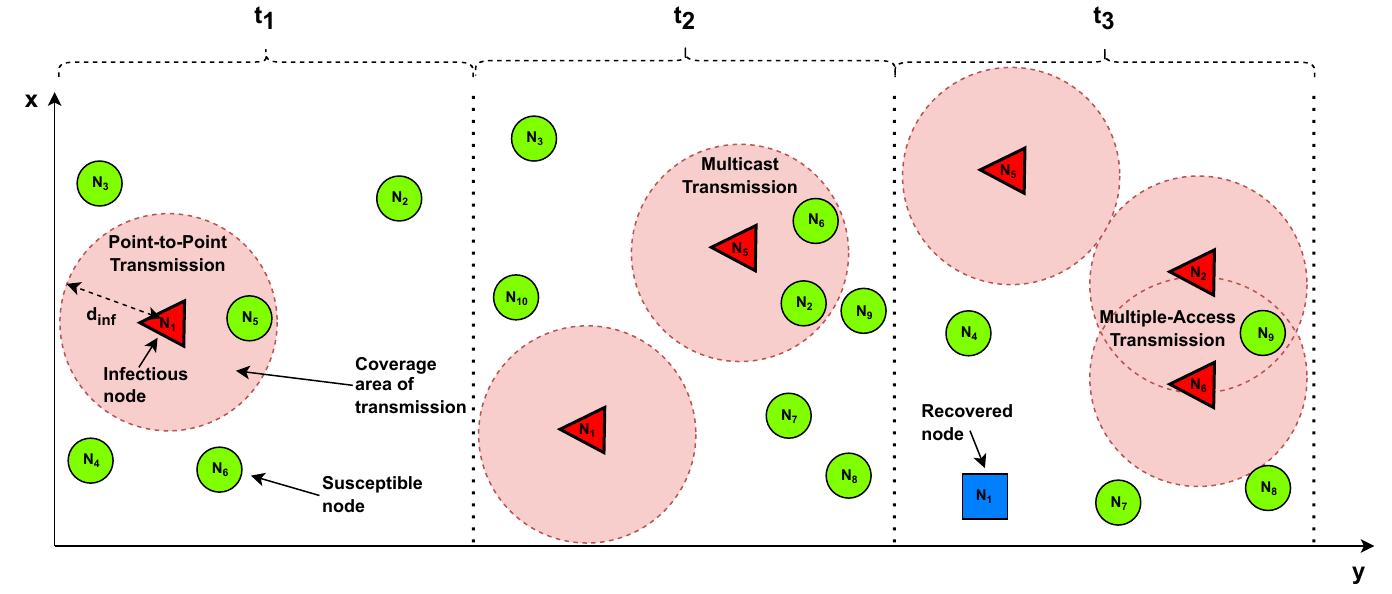}  
	\caption{
		The spread of an infectious disease in a MoHANET through 2-D space for three different time instances. As time progresses, the number of nodes changes with mobility and the nodes change their state according to their exposure to pathogen-laden droplets.
	}
	\label{MN}
	\end{figure*}

	In epidemiology literature, each human, i.e., a node, can be represented as susceptible (S), exposed (E), infectious (I) or recovered (R) according to the SEIR-based models \cite{rock2014dynamics}. Based on the disease type, different combinations of these node types can be employed for the models such as SIR or SIRS. By utilizing the widespread SIR model, a MoHANET is given in Fig. \ref{MN} which gives both the spatial and temporal changes. As the time elapses, the number of nodes may alter and nodes can make transitions between states such as S, I or R. For example, a susceptible node can become infected, if it is in the transmission range of an infectious node or an infectious node can recover after a certain period. By using a communication engineering perspective, the transmission among the nodes can be classified as given next. 
	

	\vspace{-0.1cm}
	\subsubsection{Transmission Types in MoHANETs}
	As illustrated in Fig. \ref{MN}, three transmission types are defined for the propagation of pathogen-laden droplets from the infectious nodes to the susceptible nodes as follows:
	\begin{itemize}
		\item \textbf{Point-to-Point Transmission} includes the communication between two nodes where the infectious and susceptible nodes are the TX and RX, respectively.
		\item \textbf{Multicast Transmission} is the scheme that one infectious node spreads the disease to more than one node within its communication range.
		\item \textbf{Multiple-Access Transmission} comprises the scenario where a susceptible node is exposed to pathogen-laden droplets from multiple infectious nodes.
	\end{itemize}
	
	In this paper, multicast transmission scheme is employed in the networking layer as detailed in Section \ref{OMTA}.
	
	\subsubsection{Epidemiological Models}
	The numbers of the nodes as illustrated in Fig. \ref{MN} are modeled in SEIR-based models by ordinary differential equations where these numbers can be deterministic or stochastic processes. The transition among different types of nodes are defined with certain rates which are obtained by fitting statistical epidemic data in epidemiological studies \cite{rock2014dynamics, martcheva2015introduction}. In experimental studies, these data are obtained by oral surveys or exploiting wireless sensor network technology \cite{rock2014dynamics}. On the other hand, agent-based modeling (ABM) which uses agents possessing individual contact patterns and behaviors is applied in epidemiology \cite{bissett2021agent}. Although ABM has the advantage of fine-grained results, it requires high computational power for the simulation of millions of agents, which makes it challenging \cite{bissett2021agent}. Therefore, the standard SIR model is adopted in this paper.
	
	The transition between the states in the SIR model are defined via the transition rates $\lambda_1(t)$ and $\beta_2$ as given by
	\begin{equation}
		\ce{S ->[\lambda_1(t)] I ->[\beta_2] R},
		\label{reaction}
	\end{equation}
	where $\lambda_1(t) = \beta_1 I(t)/N$, $\beta_1$ is the effective contact rate, $I(t)$ is the number of infected humans at time $t$, $N$ is the number of total population and  $ \beta_2$ is the recovery rate. The number of these node types are modeled by ordinary nonlinear differential equations as given by \cite{vynnycky2010introduction}
	\begin{align}
		\frac{d S(t)}{dt} &= \frac{-\beta_1 S(t) I(t)}{N}, \label{SIR1}\\
		\frac{d I(t)}{dt} &= \frac{\beta_1 S(t) I(t)}{N} - \beta_2 I(t), \label{SIR2}\\
		\frac{d R(t)}{dt} &= \beta_2 I(t), \label{SIR3}
	\end{align} 
	where $S(t)$ and $R(t)$ are the numbers of the susceptible and recovered humans at time $t$, respectively. Here, we assume that $N$ is constant during the epidemic so that there is not any death or birth.
	
	The solution of this equation system is generally calculated numerically. However, analytical solutions also exist as given by \cite{miller2012note}
	\begin{align}
		S(t) &= S(0) \rm{e}^{-\epsilon(t)}, \\
		R(t) &= R(0) + \frac{\beta_2 N}{\beta_1} \epsilon(t), \\
		I(t) &= N - S(t) - R(t)	
	\end{align}
	where $\epsilon(t)$ is defined as the expected number of transmissions at time $t$ and derived as
	\begin{equation}
		\epsilon(t) = \frac{\beta_1}{N} \int_{0}^{t} I(t^*) dt^*.
	\end{equation}
	
	This solution requires a numerical integration for $\epsilon(t)$ which is very similar to the solution in \cite{harko2014exact}. Furthermore, the final values of $S(t)$, $I(t)$ and $R(t)$ is derived analytically in \cite{kroger2020analytical}. This standard SIR model is employed as shown with the proof of concept study in Section \ref{POC}.

	\subsubsection{Routing and Mobility in MoHANETs} 
	Humans are susceptible to infectious diseases in indoor places. However, this is not the case that is encountered continuously. Instead, the risk to get infected is intermittent due to the mobility of humans. As people displace, their smart phones can communicate opportunistically with each other as they are in the communication range. The same type of networking is also used in many applications such as wireless sensor, vehicular, and flying ad hoc networks. These dynamically changing structures defined as mobile ad hoc networks (MANETs) enable communication using the infrastructure at their location without a dedicated router. Therefore, a MoHANET can be resembled as a specific type of MANET, that is, a delay tolerant network (DTN) in which an end-to-end link among the nodes may not always exist. The nodes in a DTN store their data and wait until they find a suitable connection. By considering this waiting delay, routing algorithms in DTNs provide the path to the desired user. Similarly, an infected human can only infect via airborne transmission once it has a susceptible human within the coverage area of transmission as shown in Fig. \ref{MN}. Hence, opportunistic routing protocols such as \textit{epidemic} or \textit{spray and wait} can be adopted to model the spread of the infectious diseases. Interestingly, \textit{epidemic routing protocol} which is a reference method for routing in MANETs was already inspired by the mechanism of infectious disease spread during an epidemic \cite{vahdat2000epidemic}. Furthermore, the mobility models employed for MANETs can be utilized to model the spread of an infectious disease in a MoHANET as given in the next part.

	\section{Proof of Concept for the MoHANET Architecture} \label{POC}
	\subsection{Omnidirectional Multicast Transmission Algorithm} \label{OMTA}
	In the SIR model, the effective contact rate ($\beta_1$) is employed to find the rate of transitions from state S to state I which is generally estimated by epidemic data \cite{vynnycky2010introduction}. $\beta_1$ is defined as the multiplication of the number of contacts per unit time ($k_e$) and the probability of infection when there is a contact between an infectious and a susceptible human ($\beta_0$) as given by \cite{vynnycky2010introduction, fang2020transmission}
	\begin{equation} \label{beta1_in}
	    \beta_1 = k_e \beta_0.
	\end{equation}
	
	By using mobility models as applied in MANETs, the average contact rate of humans ($\bar{N}_c$), which corresponds to $k_e$, can be determined via mobility models. As given in the literature about the human mobility at different scales, humans follow a Lévy walk pattern \cite{gonzalez2008understanding}. In this pattern, the flight is defined as the longest distance in a straight line between two points. During their movement, humans also stop moving for several reasons such as staying at home, or working in the office. These walks are characterized by heavy-tailed distributions such as Lévy alpha-stable (or hereafter called Lévy), lognormal or power-law for the flight lengths and pause times \cite{zhao2015explaining}.
	
	In this paper, the truncated Lévy walk (TLW) model is employed for the mobility of humans, since this model relies on real location data of humans and also has the applicability advantage of a random walk for an ad hoc network  \cite{rhee2011levy}. In TLW, each node follows a Lévy walk pattern. After each flight, a node stops for a certain pause time ($\Delta t_p$). The flight time ($\Delta t_f$) is calculated according to the relation based on the GPS traces as given by \cite{rhee2011levy}
	\begin{subequations}
		\begin{numcases}
			{\Delta t_f =}
			30.55 \hspace{1mm} \Delta r^{0.11}, &\hspace{0mm}$ \Delta r < 500$ m  \label{Delta t_f1} \\
			0.76 \hspace{1mm} \Delta r^{0.72}, &\hspace{0mm}$ \Delta r \geq 500$ m  \label{Delta t_f2},
		\end{numcases}
	\end{subequations}
	where $\Delta r$ is the flight length. Here, $\Delta r$ and $\Delta t_p$ are random variables which have Lévy distributions having the probability density function (pdf) with respect to the inverse Fourier transform of its characteristic function as given by \cite{nolan2020univariate}
	\begin{equation}
		f_X(x;\alpha,c) = \frac{1}{2\pi} \int_{-\infty}^{\infty} \exp(-jtx-|ct|^{\alpha}) dt,
	\end{equation}
	where $c$ and $\alpha$ are scale and shape parameters, respectively. In TLW, Lévy distributions $\big(S(\alpha_t, c_t) $ and $S(\alpha_r, c_r) \big)$ are truncated with an interval $0 \leq \Delta t_p < \tau_p$ and $0 \leq \Delta r < r_1$ for pause time and flight length, respectively.
	
	In addition, the average probability of infection $\bar{P}_{inf}$ can be derived by considering the propagation and reception of pathogen-laden droplets in physical and reception layers of the MoHANET. By using the system model in \cite{gulec2021molecular} whose results are given in Fig. \ref{Traj_R2}, $\bar{P}_{inf}$ is given by \cite{gulec2021molecular}
	\begin{equation} \label{P_inf}
		\bar{P}_{inf} = P(\bar{N}_R > \gamma) = Q\left(\frac{\gamma - \mu_R}{\sigma_R}  \right)	
	\end{equation}
	where $\bar{N}_R$ is the average received number of droplets with $\mathcal{N}(\mu_R, \sigma_R^2)$ and $\gamma$ is the detection threshold as defined in Section \ref{RL}. Here, $\mu_R$ and $\sigma_R$ depend on the physical layer parameters such as densities of droplets and air, dynamic viscosity of air, distance, initial velocity and size distribution of droplets and reception layer parameters such as the dimensions of the human face. Here, the infection range ($d_{inf}$) which defines the radius of the circular coverage area of airborne transmission as illustrated in Fig. \ref{MN} is introduced as a new parameter. Within this coverage area, it is assumed that $\bar{P}_{inf}$ does not change according to the distance between the TX and RX.
	
	$\bar{P}_{inf}$ given in (\ref{P_inf}) corresponds to the parameter $\beta_0$ given in (\ref{beta1_in}) in the SIR model. Thus, by using (\ref{P_inf}) and $\bar{N}_c$ computed via TLW, we propose that the effective contact rate to be employed in the SIR model can be derived by 
	\begin{equation}
		\beta_1 = \bar{N}_c \bar{P}_{inf}.
		\label{beta_1}	
	\end{equation}

It is important to note that (\ref{P_inf}) is derived for a channel with still air and a receiver with only outer reception layer. Therefore, the inclusion of different transmission mechanisms such as survival of pathogens or thermofluidic boundary conditions and also inner reception layer issues are left as open research issues. In addition, it should be underlined that (\ref{beta_1}) and (\ref{P_inf}) provide the connection between the airborne transmission and epidemiology studies by using communication engineering within the MoHANET architecture. The following describes the method of how this connection is established.
	
\begin{algorithm}[!b]
	\caption{Omnidirectional Multicast Transmission Algorithm}
	\label{Alg1} 
	\begin{algorithmic}[1]
		\State \textbf{Input:} $t_{sim}$, $\tau$, $N$, $x_{max}$, $y_{max}$, $d_{inf}$, $\alpha_r$, $c_r$, $ r_1 $, $\alpha_t$, $c_t$, $\tau_p$, $\mu_R$, $\sigma_R$, $\gamma$
		\State Determine the initial $x$ and $y$ positions of the nodes with $U(0, x_{max})$ and $U(0, y_{max})$, respectively \\
		\Comment{\textbf{Step 1: Truncated Lévy Walk}} \hbox to 0.14\textwidth{}
		\State $N_s = t_{sim}/\tau$ \Comment{Number of time steps}
		\For{$ i = 1:1:N $}
		\State $k = 1$
		\While{$k \leq N_s$}
		\State Generate $\phi \sim U(0, 2\pi)$
		\State Generate $ \Delta r \sim S(\alpha_r, c_r)$ with $0 \leq \Delta r < r_1$
		\State Generate $ \Delta t_p \sim S(\alpha_t, c_t)$ with $0 \leq \Delta t_p < \tau_p$
		\State $\Delta x = \Delta r \cos(\phi)$; $\Delta y = \Delta r \sin(\phi)$
		\State Calculate $\Delta t_f$ by (\ref{Delta t_f1})-(\ref{Delta t_f2})
		\State $n_f = \lfloor \frac{\Delta t_f}{\tau} + \frac{1}{2} \rfloor$
		\State $ x(i,k+1:k+n_f) = x(i,k) + (\frac{\Delta x}{n_f}:\frac{\Delta x}{n_f}:\Delta x)$
		\State $ y(i,k+1:k+n_f) = y(i,k) + (\frac{\Delta y}{n_f}:\frac{\Delta y}{n_f}:\Delta y) $
		\State $n_p = \lfloor \frac{\Delta t_p}{\tau} + \frac{1}{2} \rfloor$
		\State $ x(i,k+1+n_f:k+n_f+n_p) = x(i,k+n_f) $
		\State $ y(i,k+1+n_f:k+n_f+n_p) = y(i,k+n_f) $
		\If{$i^{th}$ node is out of borders}
		\State Bounce the node back from the border
		\EndIf
		\State $k = k + n_f + n_p$
		\EndWhile			
		\EndFor
		\\ \Comment{\textbf{Step 2: Omnidirectional Transmission}} \hbox to 0.06\textwidth{}
		\State $N_{c}(1,:) = 1$ \Comment{One infected node initially}
		\For{$ k = 1:1:N_s $}
		\For{$ix = 1:1:N$}
		\If{$ N_{c}(ix,k) == 1 $}
		\For{$i = 1:1:N$}
		\If{ $d < d_{inf}$ and $i \neq ix$}
		\State $N_{c}(i,k:end) = 1$ 
		\EndIf
		\EndFor
		\EndIf		
		\EndFor
		\EndFor
		\State $N_{ct} = \sum_{i=1}^{N}N_{c}(i,:) $
		\State $\bar{N}_c = \frac{1}{N_s t_{sim}} \sum_{k=1}^{N_s} N_{ct}(k) $
		\State $\bar{P}_{inf} = Q\left(\frac{\gamma - \mu_R}{\sigma_R}  \right) $ 
		\State $\beta_1 = \bar{N}_c \bar{P}_{inf}$			
	\end{algorithmic} 
\end{algorithm}

	In order to calculate the effective contact rate based on the mobility and airborne transmission, an omnidirectional multicast transmission (OMT) algorithm is proposed as given in Algorithm \ref{Alg1}. In the first step of this algorithm, the 2-D positions of the nodes in a MoHANET is determined via the aforementioned TLW. At the start of the simulation, the nodes are uniformly distributed in a 2-D rectangular area which has limits ($x_{max}$, $y_{max}$). For each flight, the nodes choose a random direction from a uniform distribution $U(0, 2\pi)$. At each flight, the positions of the nodes are checked so that they do not go outside the defined borders. If they reach the border, then the nodes bounce back to the simulation zone.

	In the second step of the OMT algorithm, the time course of the disease spread is evaluated. For each node at each time step, the contact state is checked according to the distance between the infectious and susceptible nodes. Here, it is assumed that both the infectious node emits and the susceptible node receives the pathogen-laden droplets in an omnidirectional way. Hence, if the distance between these nodes is smaller than $ d_{inf} $, then the susceptible node is assumed to be contacted and "potentially infected". The actual infection state of the population is determined according to the probability of infection, since a contact does not necessarily lead to an infection. Then, the average number of contacts of the nodes ($N_{ct}$) is calculated by summing the number of contacts of the nodes for each time step to distinguish the difference of the contact states with respect to time. Next, the average contact rate of the nodes ($\bar{N}_c$) is determined by taking the average value of $N_{ct}$ over time and dividing it to the simulation time ($t_{sim}$). Finally, the effective contact rate ($\beta_1$) is found by using (\ref{beta_1}) which is employed in the SIR model defined in (\ref{SIR1})-(\ref{SIR3}). 
	The idea behind the OMT algorithm is to find $\beta_1$ for a short period, e.g., half day, and exploit this contact pattern for long-term estimation of the infectious disease spread. In terms of epidemiological models, the proposed OMT algorithm stands between the SEIR-based and agent-based models. While contact patterns are determined similar to ABM, the long term time course of the epidemic is estimated by using a SEIR-based model. Next, numerical results for the OMT algorithm are given.

	\begin{table}[!b]
		\centering
			\begin{tabular}{ll|ll}
				\toprule
				\textbf{Parameter}	& \textbf{Value} & \textbf{Parameter}	& \textbf{Value}\\
				\midrule
				$t_{sim}$  &  $ 12$ h & $\tau$ & $10$ s \\
				$N$ & $ 1000 $ & $d_{inf}$ & $1$ m \\
				$x_{max}$ & $2000$ m & $y_{max}$ & $2000$ m \\
				$\alpha_r$ & $1.6$ & $c_r$ & $10$ \\
				$\alpha_t$ & $0.8$ & $c_t$ & $1$ \\
				$r_1$ & $1000$ m & $ \tau_p $ & $1000$ s \\
				$\mu_R$ & $120$ & $\sigma_R$ & $10$ \\
				$\gamma$ & $140$ & $\beta_2$ & $0.037$ \\
				$I(0)$ & $ 1 $ & $ S(0) $ & $ 999 $ \\
				\bottomrule
		\end{tabular}
		\caption{Simulation parameters}
		\label{Sim_parameters}
	\end{table}

	\subsection{Numerical Results}
	In this part, numerical results are obtained by employing the SIR model which uses the effective contact rate via the proposed OMT algorithm. The simulation parameters are given in Table \ref{Sim_parameters}. In the OMT algorithm, $\bar{N}_c$ is determined via Monte Carlo simulation for a time period ($t_{sim}$) of half day assuming that humans stay at their homes for the rest of the day. The values of the shape and scale parameters for flight and pause time distributions in TLW, i.e., $\alpha_r$, $c_r$, $\alpha_t$, $c_t$, and their truncation values, i.e., $r_1$, $\tau_p$, are chosen according to values based on real location data in \cite{rhee2011levy} for a $2000$ m $\times$ $2000$ m area. In the simulation of the SIR model, it is initially assumed that there is only one infected human in the MoHANET, other nodes are susceptible. In addition, it is assumed that the number of people does not change due to death, birth, etc. during the simulation due to the constant population density of Italy \cite{worldometer2020update}. The idea of the OMT algorithm is to find an average contact rate for a  limited population in a limited area with a realistic mobility pattern and to apply it by scaling to a larger area and population. The parameters of physical and reception layers such as $\mu_R$, $\sigma_R$, $\gamma$ and $d_{inf}$ are determined in accordance with the values in \cite{gulec2021molecular}. In addition, please note that $\beta_1$ value is converted from the unit $s^{-1}$ to $day^{-1}$ to be used in the SIR simulations.

	\begin{figure}[t]
		\centering
		\includegraphics[width=\columnwidth]{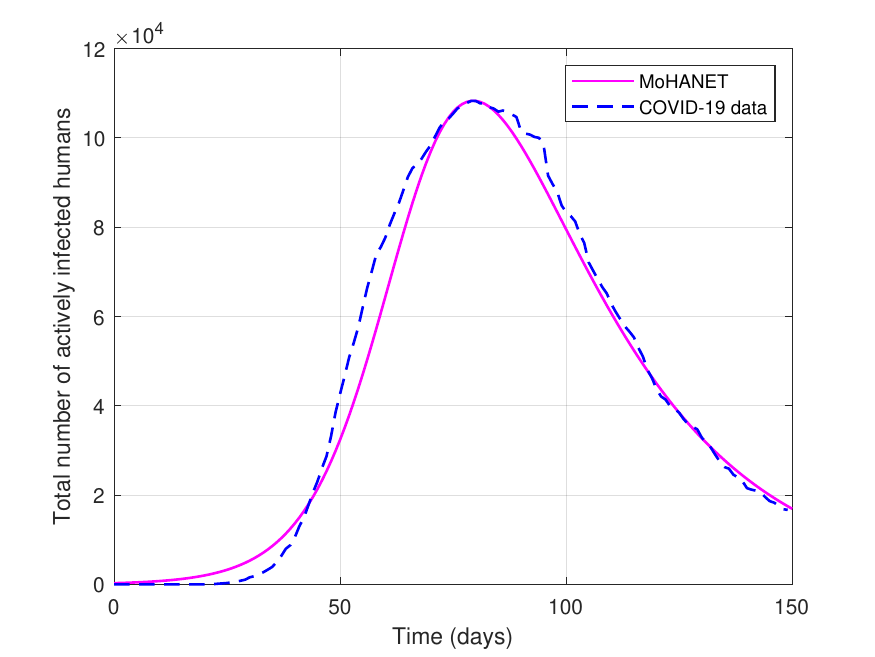}
		\caption{
			Number of infected humans (nodes) in a MoHANET with $\gamma = 142.4$, $\beta_2 = 0.037$ and the COVID-19 data of Italy for the first 150 days (31 January 2020-28 June 2020) at the beginning of the pandemic. The root mean square error between these curves is $5110.2$.
		}
		\label{SIR}
	\end{figure}
	
    \begin{figure}[b]
    	\centering
    	\includegraphics[width=\columnwidth]{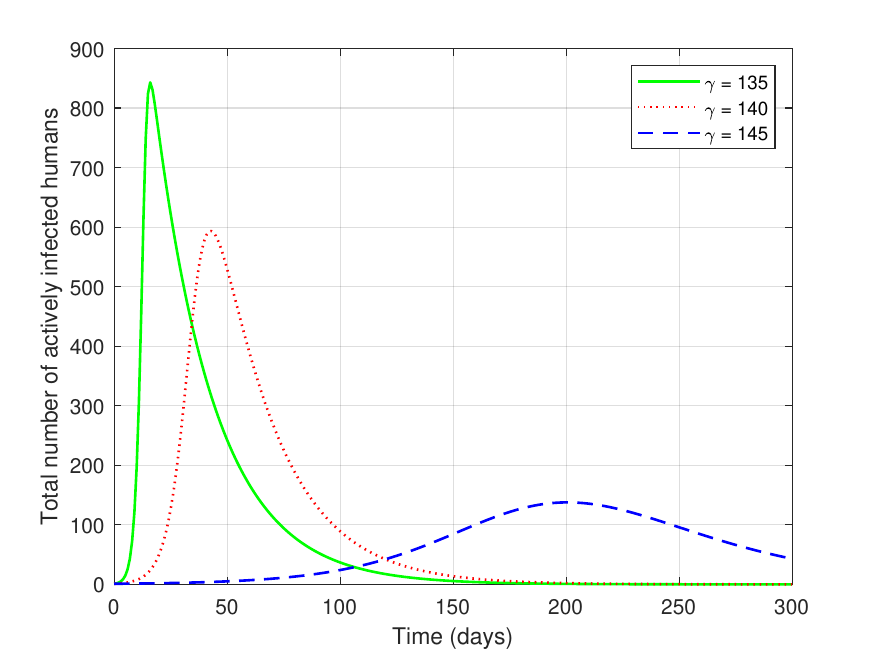} 
    	\caption{
    		Total number of actively infected humans (nodes) in a MoHANET for different threshold values. 
    	}
    	\label{gamma}
	\end{figure}

	In Fig. \ref{SIR}, the results obtained with the MoHANET architecture are validated by the active infected cases of the COVID-19 outbreak in Italy for the first $150$ days (31 January 2020-28 June 2020) by using the data set in \cite{dong2020interactive}. The start date of the data is the first day an active case was recorded. In this data set, the number of actively infected cases $\left(I_d(t)\right)$ are calculated by subtracting the number of recovered and death humans from the number of total confirmed cases. The number of deaths are omitted in the SIR model, since it is negligible with respect to the population of Italy ($\approxeq 6 \times 10^7$). In addition, the estimated number of infectious humans is scaled via multiplying $I(t)$ by $\max(I_d(t))/\max(I(t))$ where $\max(.)$ shows the peak value, as also applied in \cite{cooper2020sir}. In order to visually fit the COVID-19 data, $\gamma$ and $\beta_2$ values, which are given in the caption of Fig. \ref{SIR}, are arranged manually in order to fit the empirical data visually and having a root mean square error as small as possible. The chosen value of $\beta_2$ is also in agreement with a similar SIR modeling study for COVID-19 \cite{cooper2020sir}. This figure shows that airborne transmission issues and mobility can be combined with the SIR model within the MoHANET architecture. Here, $\bar{N}_R$ depends on the parameters such as the velocity and size distribution of droplets, air-droplet interaction and receiver geometry. Thus, this modeling approach gives the opportunity to include the parameters of the physical and reception layers in the networking layer of the MoHANET. For convenience, $\beta_2$ for the transitions from state I to state R is taken as a constant value. However, $\beta_2$ can be estimated at the reception layer by using human's immune system response or drug-human interaction at the cellular level. 
	
\begin{figure}
		\centering
		\includegraphics[width=\columnwidth]{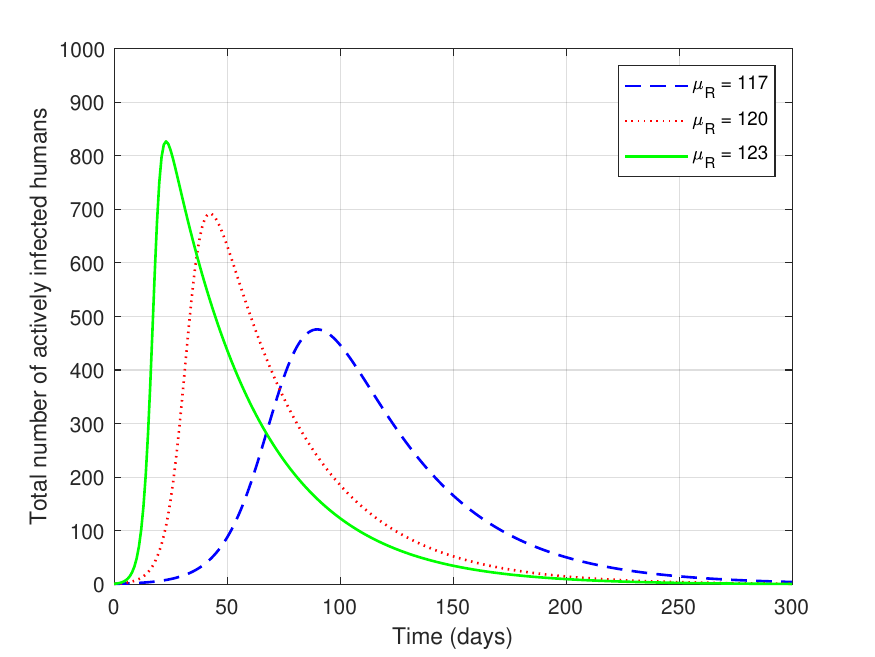} 
		\caption{
			Effect of the received number of droplets on infectious disease spread through airborne transmission.
		}
		\label{mu_R}
	\end{figure}

\begin{figure}
	\centering
	\includegraphics[width=\columnwidth]{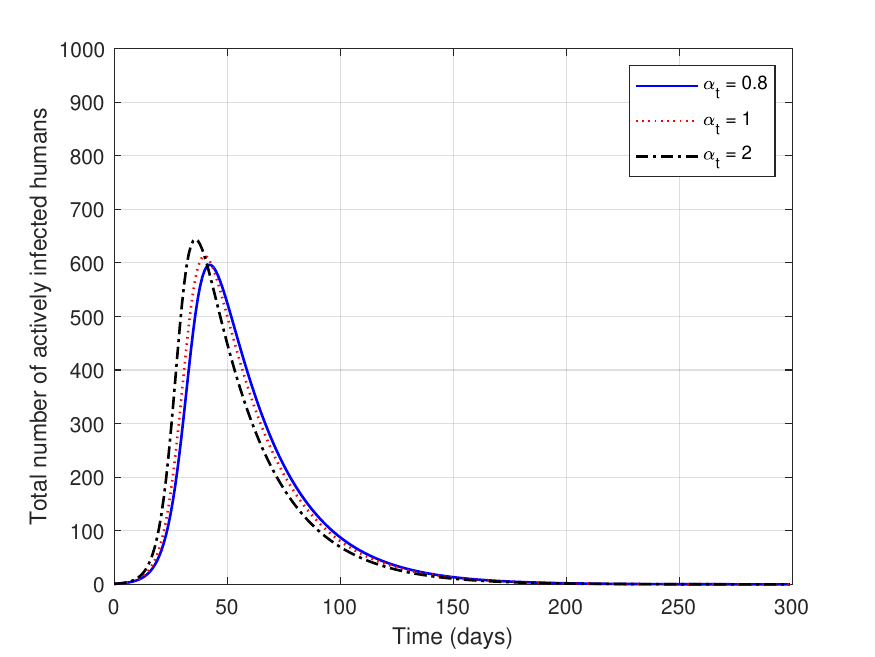} 
	\caption{
		Effect of different pause time distributions on infectious disease spread through airborne transmission for $d_{inf}=1$ m and $\gamma = 140$.
	}
	\label{alpha_t}
	\end{figure}

	Surely, the empirical data set includes situations such as people with and without masks or limited mobility due to lockdown which lasted from 9 March 2020 until 18 May 2020 in Italy. Furthermore, the population is not  stationary even in a lockdown. However, the results obtained with the MoHANET architecture aim to give an average estimate for an epidemic as also applied in the epidemiology literature with the SIR based models. The effects of different scenarios are included in the proposed model via arranging the parameters in (\ref{P_inf}) for airborne transmission issues and TLW parameters for mobility issues as also given in Figs. \ref{mu_R} and \ref{alpha_t}. In the rest of this section, the results are obtained for the parameters given in Table \ref{Sim_parameters} without scaling and for a generic scenario to show the effects of the different layers of the MoHANET on an epidemic.

	The results given in Fig. \ref{gamma} depicts the effect of reception layer issues on the time course of the epidemic. Here, it is shown that the rate of an epidemic reduces as $\gamma$ which depends on the average strength of the humans' immune system increases. This actually corresponds to the fact that people in a society can reduce the impact of epidemics by preferring a healthier lifestyle that includes practices such as following a balanced diet, living in a less stressful way or exercising regularly. Here, it can be observed that a small change in $\gamma$ can affect the course of the epidemic dramatically. This shows that health authorities should promote and support to boost the immune system strength of the population constantly as well as supporting vaccine development for infectious diseases.

	
	In Fig. \ref{mu_R}, the effect of the mean received number of droplets ($\mu_R$) during the airborne pathogen transmission on an epidemic is shown. In case of a longer stay in an indoor place, a more violent respiratory event, a closer distance to an infectious human or not wearing a mask, $\mu_R$ of a susceptible human can increase. Hence, Fig. \ref{mu_R} can be interpreted as the effect of social distance rules such as wearing masks in an epidemic. If these rules are followed, i.e., $ \mu_R $ is decreased, the curve can be flattened in an epidemic. In addition, Fig. \ref{mu_R} can also be considered as the effect of different types of masks. As the filtering capacity of the mask used in the population increases, e.g., FFP-2 mask, the epidemic has a milder course. On the other hand, the increased infectivity of a pathogen along the different stages of an epidemic due to its evolution can be simulated by increasing $ \mu_R $ as shown in Fig. \ref{mu_R}.
	

	Fig. \ref{alpha_t} depicts the time course of the infectious disease spread for different pause time distributions of the nodes. As $\alpha_t$ decreases, the nodes in the MoHANET have shorter pause times at their locations. Hence, their contact rate decreases. This also corresponds to the situation where people avoid long visits to each other and less people get infected. Here, $\alpha_t=2$ corresponds to a normal distribution which severely differs from TLW with the realistic value $\alpha_t=0.8$. Thus, these results show that the mobility parameters should be chosen carefully according to the scenario so that they affect the estimations of the time course of an epidemic. In addition, these results can also be considered as the effect of a lockdown in an epidemic. As $\alpha_t$ decreases, the degree of mobility decreases. Hence, it is shown that the lockdown can help to decrease the number of infected people. For realistic results on the effect of airborne transmission, several experimental methods can be employed, which is discussed in the next section.

	\section{A Discussion on Experimental Techniques and Simulations for MoHANETs}
	In order to observe and model the airborne transmission mechanisms among humans, experimental setups and computer simulations can be employed. In this section, we present and discuss how the performance of the proposed methods in different layers of the MoHANET architecture can be evaluated. 
	
	In physical and reception layers, the emulation of breathing, coughing and sneezing in experimental setups are realized by respiratory machines or thermal manikins which can be heated to change their temperature. These devices emit tracer gases including droplets. The concentration of droplets is measured by air samplers or via imaging techniques such as particle image velocimetry which gives the velocity and directions of droplets \cite{ai2018airborne}. Moreover, sprayer-based MC systems can also be used instead of respiratory machines, manikins and air samplers \cite{gulec2021fluid, gulec2020droplet}.
	
	Albeit reliable results can be obtained by physical experiments regarding the consideration of droplet-air interaction and airflows, collected data have a low-resolution in space and time and experimental devices are expensive. Therefore, computational fluid dynamics (CFD) simulations are employed to evaluate the airborne transmission mechanisms with a high spatiotemporal resolution and less cost \cite{ai2018airborne}. However, the simulation software programs are based on Navier-Stokes equations which lack the capability to model all of the effects during the transmission realistically.
	
	These experimental techniques and CFD simulations can be employed in physical and reception layers \cite{vuorinen2020modelling, pendar2020numerical}. In the networking layer, it is essential to model the spread of infectious diseases with an approach that takes into account the interaction of people and their mobility in both time and space. The movement patterns of humans can be simulated by synthetic models or trace-based models which rely on real mobility data of mobile nodes \cite{solmaz2019survey}. The adapted routing protocols for MoHANETs can also be evaluated in time and space by employing these mobility models according to the scenario via network simulation software. With a holistic perspective, new software is needed to model all of the issues at different layers of the MoHANET in a single platform.
	
	\section{Conclusion}
	This paper presents a framework to model airborne pathogen transmission with a communication engineering perspective. First, airborne pathogen transmission mechanisms are reviewed and MC is utilized to model the propagation and reception of this transmission. The concept of MoHANET is proposed to handle the infectious disease spread modeling problem by using a layered structure in macro- and microscales. Furthermore, a proof-of-concept study is given about the MoHANET architecture via the proposed OMT algorithm. The holistic viewpoint of communication engineering can bring different disciplines such as fluid dynamics, medicine, biology  and epidemiology together for accurate predictions about the spread of infectious diseases. Numerical results confirm that prevention measures that reduce the number of pathogen-laden droplets received by individuals or a healthier population, i.e., people with stronger immune systems, can flatten the curve in an epidemic.
	
	\section*{Acknowledgment}
	The work of Fatih Gulec was supported in part by the Scientific and Technological Research Council of Turkey (TUBITAK) under Grant 119E041 and in part by German Academic Exchange Service (DAAD). The work of Baris Atakan was supported by the Scientific and Technological Research Council of Turkey (TUBITAK) under Grant 119E041. The work of Falko Dressler was supported by the project MAMOKO funded by the German Federal Ministry of Education and Research (BMBF) under Grant 16KIS0917.
	
	
\bibliography{ref_fg_MoHANET}
\bibliographystyle{elsarticle-num}


	%

\end{document}